# The Evolution of Language and Human Rationality

Robert Worden

Theoretical Neurobiology Group, University College London, London, United Kingdom

[rpworden@me.com](mailto:rpworden@me.com)

May 2024

Abstract:

This paper distinguishes between narrow language (syntax and semantics) and full language, as it permeates our lives. Full language includes dialogue pragmatics, the human Theory of Mind, self-esteem and social emotions. Narrow language and full language both depend on a fast, powerful, pre-conscious Bayesian pattern-matching process of feature structure unification.

Language evolved up to two million years ago, partly by sexual selection for the display of superior intelligence. Sexual selection accounts for the uniqueness of human language: other species have not faced the same sexual selection pressures. It also accounts for the prodigious speed and expressivity of language, which goes far beyond what we need in natural habitats; but we use prodigious language to show off superior intelligence. Sexual selection is not always for the best; it is accompanied by handicaps, such as the peacock's tail, or our enlarged, expensive brains.

If language is used to display intelligence and compete for mates, it must be accompanied by other abilities: a Fast Theory of Mind (to converse) and social emotions (to seek high status within a group, to find a mate). Language is the result of evolution, not enlightenment. Narrow language, the vehicle of our rational thought, is part of full language, which brings with it a less rational side of human nature – our irrational emotions, and the harm we often do to ourselves and others. It is an urgent scientific priority to understand the irrational side of human nature; full language is the starting point for doing so.

## 1. Introduction

*Le cœur a ses raisons que la raison ne connaît point - Pascal*

The human mind is a prodigious pattern-matching engine. Throughout our lives, we learn thousands of patterns, and we rapidly retrieve them to match them to whatever we are experiencing. Language is built on this pattern-matching ability – matching the sounds of words, to understand what we hear. Language is generally seen as a benign, neutral medium for creating and expressing ideas – a wholly beneficial adaptation of the human mind. This view of language is shown in figure (1a).

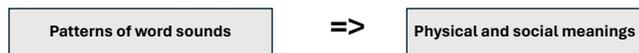

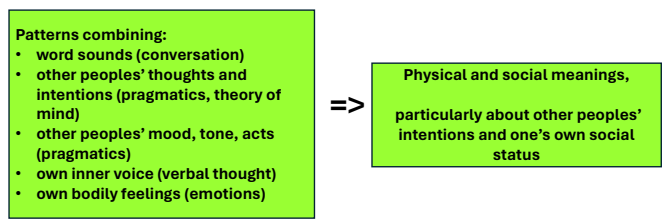

*Figure 1: a narrow view of language, and a more complete view of language*

The narrow 'syntactic/semantic' view of language, as simply a relation between words and meanings, derives from the academic study of syntax and semantics – which are the intellectually purer aspects of language. Looking at the many uses of language, and the many ways it influences our lives, suggests a more complete view. Language is more pervasive, and language pattern matching is more wide-ranging, encompassing all the patterns of figure 1(b). Language is not just the benign, neutral medium we have taken it for; its role in our minds is more profound, and not always beneficial. As well as being a neutral vehicle for thought, it is a pre-conscious control mechanism for much of our lives. The reasons for this lie in the evolutionary origins of language – which involve sexual selection. Understanding these origins can help us understand the role language plays in our lives, and how it is linked to a less rational side of our nature.

## 2. From Narrow Language to Full Language

Rational choices are conscious choices. To make a rational choice, we consciously consider a number of options, and pick the one which we regard as the best. A choice cannot be called rational, if alternatives have not been considered.

Human thought cannot be rational 'all the way down'. If we had to think how to think, we would never start thinking. There has to be some innate base level of cognition, which is trusted to be the basis of rational thought. That base level is commonly taken to be language – which is seen as the medium of rational thought. Our basic ability to use words is not thought of as rational. Language semantics and syntax are pre-conscious. Nobody has ever made the conscious choice to call a barking quadruped a 'dog'. For the word 'give', nobody ever consciously chose to precede it by the giver, then follow it by the recipient, and finally by the gift. These usages developed in society by a consensus over many generations, through no conscious choices, and are passed on by learning to every child [Worden 2002]. In themselves these choices are not rational, but we trust them to express rational thoughts. We create the thoughts by a fast, automatic, pre-conscious process of creating words; but we critique them rationally.

So narrow language – the syntax and semantics of words – is not itself a rational process; but the thinking we do with it is (we believe) rational.

This paper argues that the same fast, pre-conscious pattern matching process[1] which drives the narrow language of figure (1a) also drives the full language of figure (1b); that full language controls much of our daily behaviour – how we regard our own lives and our place in society; how we understand other people; our motivations, and how we feel emotions. These aspects of full language are (like narrow language) fast, pre-conscious and automatic. They are not rational – they are not the result of consciously weighing alternatives. The consequences of this for humanity have been both good and bad.

To see how full language controls our lives, we need to understand how human language evolved, and how it works computationally in the brain. These are addressed next.

## 3. Language Evolution and Sexual Selection

There are many theories about how human language evolved, described in previous proceedings of this conference, and in (Christiansen & Kirby, eds, 2003). These theories face two difficulties (Szamado & Szamasthary 2006):

A. They do not account for the uniqueness of human language. If mankind has expressive language and high intelligence, why has no other species evolved a similar capability?

B. In most accounts, the fitness benefits brought by language in a natural habitat are not

---

[1] technically, this is the unification of feature structures, or of language constructions

sufficient to offset the large metabolic costs of our greatly expanded brains.

An account of language evolution through both sexual selection and natural selection can address these problems.

Sexual selection (Lande 1981; Maynard Smith 1982) is very widespread. It creates much of the diversity and vivid profusion of nature, such as birds' plumage or flowering plants. (Worden 2022) has proposed a hybrid account of the evolution of language, in which both natural selection and sexual selection have played a part. In this account, superior intelligence became a sexually attractive trait in *Homo Sapiens*, needed by both sexes to attract a mate (Miller 2002); and complex language evolved as the primary way to display superior intelligence. This hybrid account does not conflict with accounts of language evolution by natural selection. For a full account, see (Worden 2022). In short, both of the difficulties (A) and (B) are addressed by sexual selection:

   A. Sexual selection leads to species-unique traits, because it acts in a unique way within each species; so other species do not have language
   B. Sexual selection is a process of runaway positive feedback, leading to exaggerated traits and handicaps, such as the peacock's tail, or the metabolically expensive human brain. So we have super-powerful language. It does not maximise the survival fitness of individuals in a natural habitat.

If language evolved for the display of intelligence, to be sexually attractive and to gain high social status (in order to get a mate), some key properties of language follow:

   1. Language must be accompanied by high general intelligence, in order to be impressive (our enlarged brains)
   2. Intelligence is displayed through conversation; the skills of conversation are a key part of language (pragmatics)
   3. To impress, our speech must be fast and expressive (complex syntax, large vocabulary)
   4. To impress another person in conversation, you need to know what they think, know and do not know (this requires a Theory of Mind, or ToM)
   5. You need to read their intentions though their gestures, tone of voice, and facial expressions – as well as their words.
   6. To gain high status (in other peoples' eyes), requires inferring what they are thinking about us
   7. Our concept of ourselves and our status is defined by what we think other people think about us (self-esteem)
   8. To make our conversations more impressive, we rehearse them internally (verbal thought)
   9. We monitor our changing self-esteem through our bodily feelings (emotions)

This helps to understand all the fast pattern-matching we use in language - the full language of figure 1(b), not just the narrow language of figure 1(a). It shows that language is deeply linked to our self-esteem and emotions.

## 4. Computational Models of Language

The pattern-matching which narrow language (syntax and semantics) depends on, is not just some fuzzy, ill-defined form of pattern matching. It is a precise and powerful computable form of pattern matching (called **unification**) which has been studied for many years, and underpins the syntactic structures of all languages.

The computational study of language has seen a long-running schism between followers of Noam Chomsky[2] (1965, 1981, 1995) and others who have not followed him. Initially, these others built unification-based computational models of language (e.g. Kaplan & Bresnan 1981), and later united under the banners of cognitive linguistics and construction-based grammar (Fillmore 1985, Kay 2002, Goldberg 1995, Langacker 2008, Sag, Boas & Kay 2012).

Most of the work of building computational models of language has been done by the latter school. It has led to a detailed working computational model of how language is processed in the brain – which is supported by its agreement with data on many diverse languages.

I summarise here the computational model of cognitive linguistics, to bring out its implications for the learning and real-time processing of language:

   1. Words and other language constructions are represented in the mind as **feature structures**. A feature structure is a tree-like structure of nodes with properties on each node, which represents both the sounds in a word, and its meaning.
   2. The primary operation in language production or understanding is **unification** – a well-defined mathematical form of pattern matching and pattern merging, in which a construction is matched with sounds heard (for understanding) or intended meaning structure (for production), expanding a feature structure by adding new structure – adding meaning for language understanding, or adding words sounds for language production.
3. This computational model of language processing works well for all known languages, explaining how meanings in the brain are converted to word sounds (for production) and how word sounds are converted to meanings (for understanding).
4. The unification model has been combined with Bayesian probability, showing that unification is a form of Bayesian maximum likelihood pattern-matching, able to match patterns and find the most likely meaning in the presence of many kinds of noise, uncertainty and ambiguity.
5. There is a second Bayesian operation on feature structures which is complementary to unification, called **generalization**. The generalization of two feature structures results in a third feature structure, simpler than the two input structures, containing only their shared common structure.
6. Feature structure generalisation is the basis of a working computational model of language learning, which can learn the syntax and semantics of a language (Worden 1997, 2022b), in a way that mirrors how children learn their native language – learning any word or other construction from hearing only a few examples of its use, inferring the meaning of learning examples from other cues.

---

[2] pursuing an evolving succession of theories - from deep structure, to transformational grammar, to movement and binding, to the minimalist program.

Human language rests not on some loose, approximate form of pattern matching – but on the fast, powerful, and well-defined computation of unification, which matches the patterns of several words per second when producing or hearing an utterance. Unification takes place outside our conscious awareness, before we become aware of the meaning of any utterance (for instance by seeing a mental image of a meaning). We learn the constructions for thousands of words by the unconscious precise process of feature structure generalization; we retrieve and use those feature structures with no conscious effort.

Why are our minds capable of this prodigious, precise pre-conscious computation? It has evolved by natural and sexual selection, driven by the strong selection pressure to use language to show off intelligence, to get high status in the group and to get a mate (Worden 2022a). To impress others, our use of language needs to be prodigious – deploying thousands of words in complex utterances without hesitation or effort. Sexual selection for language is an intense competition. People whose language was slower, less expressive or hesitant failed to impress others, and failed to get a mate.

The narrow core of language (syntax and semantics) relies on the fast, precise and powerful operations of unification and generalization. This paper proposes that the same fast operations underpin all the skills of full language, shown in figure 1b.

## 5. Pragmatics and the Theory of Mind

The first use of language is in conversation. To impress other people with our intelligence, we need to be fluent conversationalists – able to take our conversational turns within a fraction of a second (Levinson & Torreira 2015), to rapidly infer the relevance to the conversation of what someone has just said (Sperber & Wilson 1986), and to infer our partner's conversational intent from what they say, and from the context. These pragmatic skills require mind-reading – a Theory of Mind (ToM), to infer what the other person in a conversation is thinking (Sperber & Wilson 2002); so that a shared cooperative intent in the conversation is part of the common ground (Stalnacker 2002, Tomasello 2014) shared between speaker and listener.

Because conversational skills need to be deployed very fast – to take our conversational turns within a fraction of a second – they cannot depend on conscious deliberation, such as mental imagery or simulation, because that would take too long. This paper proposes that the conversational skills of pragmatics and the human theory of mind depend (like syntax and semantics) on fast, pre-conscious unification of feature structures.

Just as we learn feature structures for the meanings of words, and unify them within a fraction of a second, we also learn pragmatic feature structures for other people's states of mind and conversational intentions. We unify those feature structures rapidly and pre-consciously, so we can take our conversational turns without hesitation.

Pragmatic conversational feature structures are, like the feature structures for words, nested trees of nodes, with attributes on the nodes describing sounds and meanings. Because they represent other peoples' states of mind and intentions, they are more deeply nested trees than the feature structures for simple words. A pragmatic feature structure has nodes representing 'X thinks that ..' or 'X wants that…', with nested nodes representing what X thinks or wants. These feature structures are deeper and more complex, but we still unify them pre-consciously, very fast and with no conscious effort; we still learn them automatically by generalization, from examples of conversations. They are the basis of our conversational skills.

So the human Theory of Mind – which evolved through our need to be fast and fluent conversationalists, in order to get a mate – does not work by rationally, deliberately considering what another person is thinking or planning or wants, considering and rejecting possibilities. It works by fast, pre-conscious pattern matching – a **Fast Theory of Mind (FToM)** – using patterns we have learnt in previous conversations. The Theory of Mind is not a rational conscious process.

This is a first sense in which the full language of figure (1b) relies on feature structures and unification; and in which language is not as rational as it is often taken to be.

Our mind-reading in conversations is fast, pre-conscious, and sometimes unreliable. Just as we can misunderstand or mis-hear words or utterances, we can misunderstand the intentions of a conversational partner. In conversation there are other cues such as a speaker's tone of voice or gestures, which we use for pattern-matching; and if we misunderstand someone, they can correct us, to repair the conversation. The unreliability of the human fast theory of mind is not a big handicap in conversation. Outside conversations, in our private thoughts, it is more of a problem.

## 6. Self-Esteem and Emotion

In conversations we learn a Fast Theory of Mind – an ability to infer rapidly what a conversational partner is thinking, feeling and intending, from what they say and from contextual cues.

Crucially, what we infer includes what they are thinking about ourselves. We need to know what they think about us, in order to carry on the conversation. This is why the human sense of 'myself' is much more developed than the sense of self in many other animals. The human sense of self is largely a sense of 'what I think some other person is thinking about me'. This becomes the self as measured against the social norms of the group – seeing oneself in the mirror of other peoples' assessments (or what one believes are other peoples' assessments).

The need to impress others is linked to a need to obtain high social status within a group, in order to get a mate. Our self-perceived social status is a ToM assessment of 'what I think other people think of me'. In conversation, we track that assessment, and choose what we say next to bolster it. This is our self-esteem. Part of this process is to feel unpleasant emotions – bodily feelings that are triggered when our self-esteem is threatened - and to use those emotional feelings to guide what we say, to bolster our self-esteem when necessary. This requires us to learn the patterns of our bodily feelings arising from emotions, and to use them to guide our conversation.

This is a second sense in which the capacity for full language relies on the same fast, pre-conscious pattern matching as syntax and semantics; and in which it is less rational than we like to think. The patterns we match pre-consciously include not only word sounds and cues such as tone of voice, but also the patterns of our own bodily feelings.

Full language, expressed though self-esteem and emotion, can be inaccurate and irrational in two ways:

- The human sense of self is based on 'what I think other people think about me' – a second-hand view of oneself through the eyes of other people, using a Fast Theory of Mind which is erratic and error-prone.
- We interpret our own bodily feelings not just as physical feelings in some part of our body (such as the face, or our heart rate) but as learnt codes for some social situation, telling us that we should respond in some way. These codes are learnt by an unreliable process at a young age, when we are prone to misunderstand.

As an example of how a pattern of bodily feelings can be learnt, consider a child who is reprimanded for some naughtiness. The child knows she has done wrong (that she has violated some social code) and adopts a facial expression expressing shame. This expression is accompanied by subtle feelings in the face, from the use of certain facial muscles; and other bodily feelings. From then on, those facial and other feelings are associated with the social emotion of shame, and with low self-esteem. But the learning examples may be misleading or misunderstood; the shame may be misplaced. Like a learnt word, it is hard to unlearn it. Because the feelings are internal, unlike words, they are not subject to external correction. Emotions are driven by a fast, pre-conscious, irrational assessment of our bodily feelings.

In summary, full language (figure 1b) requires us to learn thousands of complex patterns, involving word sounds, the inferred mental states of others, and our own bodily feelings. These are shown in figure 2.

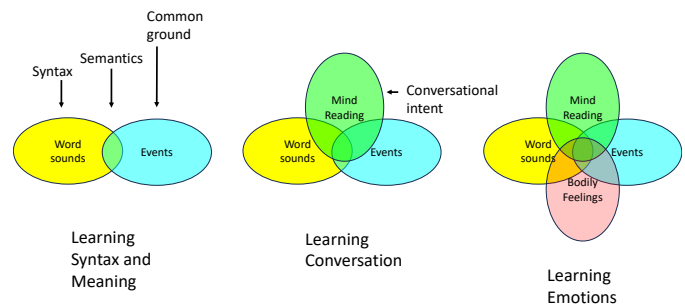

*Figure 2: Three components of language learning, which are needed to use language to display intelligence.*

As early as three years of age, a child has learnt many complex patterns, in all four quadrants of the picture (Bloom et al. 1993; Fletcher & McWhinney 1996). Human speech and thought works by fast, pre-conscious matching of these patterns, through unification. That makes the human mind a very complex and powerful dynamical system – and not always a rational one. The human mind is a product of evolution, not enlightenment.

## 7. Pragmatics as Language Games

There is a close identification between conversational pragmatics and language games, as identified by (Wittgenstein 1958).

To describe this identification (or close analogy):

- Conversational pragmatics is like a board game, in which each player makes a move in their conversational turn.
- The board is the common ground, of the players' mutual understanding of the current situation (including of each others' mental states)
- The pieces on the board are feature structures (mental representations) describing the common ground, including the players' mental states (knowledge, emotions, and intentions).
- Each player has a stock of pieces (feature structures) which they can play in any turn.
- To play a piece is to unify the feature structure (pattern-match it) with feature structures on the board, and to make an utterance which adds (by unification) to the pieces on the board, extending the common ground.
- Players acquire the pieces they need to play any language game by learning – which is done by generalizing the feature structures they have experienced when previously playing the same game.
- Games are language-specific and culture-specific; we have evolved the capacity to learn and to play many different games.
- People judge our social status by our ability to play the language games of the group – by the fluency and skill of the moves we make.
- Games are cooperative, in that neither player wishes to be seen to be responsible for the breakdown of a game.
- Language evolution is the evolution of our prodigious ability to play these games, under the intense pressure of sexual selection.

It is a familiar idea (and well established) that the syntax of an utterance is defined by unification of feature structures. This paper proposes a small extension of that idea – that the pragmatic structure of a conversation is also defined by unifying feature structures (that we have learned in previous conversations). I am currently building a computational model of pragmatic language games, on these lines.

I illustrate the points above with one of the simplest language games – the Greeting Game:

> A: Hello
>
> B: Hello

In terms of narrow language (syntax and semantics) this game is trivial. In terms of full language, it is not simple. There are pre-conditions for playing the game, which must be in the common ground between two people:

- A and B must both be present in the same physical space, for some extended time
- Neither A nor B must be fully occupied with some other activity, such as eating, sleeping, skiing, or talking to a third person.
- A and B must have roughly comparable social status within the group.
- Either A and B have previously met, or there must be some mutually understood reason for them to need to cooperate.
- A and B should not be enemies

Both A and B unify their learned feature structures with feature structures in the common ground, to determine whether these conditions are satisfied - whether or not the game can be played.

The feature structures we learn constitute our knowledge of the game in two senses: They are the rules of the game (a valid game unifies with those feature structures), and they are the skills of playing (to make a move, you unify a feature structure).

There are group-specific rules about which of A or B may initiate the greeting game, and how they may do so (allowed gestures, tone of voice, etc.). There are group-specific rules for how B may respond or not respond, and for the social/emotional impact of various outcomes.

For instance, if B does not respond in the manner expected by A, both A and B may infer things about their social status and the emotions they are entitled to feel. Both may infer that some third person C, observing the exchange, makes certain inferences about A and B.

If B fails to respond, C might infer that A is entitled to feel offended; or alternatively, that B's social status is so elevated that B need not deign to respond; or that B is just stupid. Both A and B infer that C will infer these things, and that their social status (in the eyes of the group) will change accordingly. These are Fast Theory of Mind inferences.

On its own, the greeting game has no practical purpose. It may serve as a preliminary to some practical, survival-enhancing activity like gathering food; but equally it may not. Its real purpose is that each person needs to play the game skillfully, in order to show off their high intelligence, gain high social status, and get a mate. It is part of the social control system for the group.

Frequently the greeting game leads on to other games, such as various request/response games, or to persuasion games (Mercier & Sperber 2017) – where one person tries to persuade the other person that some propositions are true. This paper is a persuasion game. There is a huge variety of language games, and they all require proficiency in both narrow language and full language.

We are fascinated with games from an early age, because our minds have evolved to do just that - to learn and play skillful language games, to avoid the severe evolutionary penalty of not passing on our genes.

## 8. Language, Thought and Rationality

Language – the medium in which rational thoughts are expressed – is built on fast pre-conscious processes, which are learnt and which are not rational. The irrational runs faster than the rational; and sometimes, the rational cannot catch up.

To make our conversations fluent and impressive, we mentally rehearse them. This is the origin of verbal thought. As we think, we are consciously aware of the sounds of the words, and we remember them. Later recall enables us to construct extended chains of thought, which is seen as the basis of our rationality (Pinker 2021).

When rehearsing conversations as verbal thoughts, we have in mind who the audience might be. As we think, the FToM patterns that we have learnt are matched, and we infer what the audience will think – their reactions to our words. In much verbal thought, there is a **Shadow Audience** in our minds, and we constantly infer what they will think about what we are thinking and might say. The shadow audience may be a specific person, but often it is an ill-defined group, such as 'my parents' or 'the neighbours' or 'my peers at work', or even just 'people'.

The influence of the shadow audience on our thought is pervasive and often irrational:

1. The many ToM patterns which we learn in conversations match sense data (what the other person says, contextual cues). Those patterns work well enough to sustain a conversation (Levinson 1983). When the same ToM patterns are matched in our private thoughts, there is no feedback from another person – no correction; so as a guide to what other people think, the ToM patterns can be wrong.
2. Much of what we infer is about ourselves: 'what my shadow audience thinks of me'; if that is negative, lower self-esteem triggers negative emotions. These are consciously felt in the body, leading to unpredictable cascades of further thoughts and emotions.

So while verbal thought enables us to construct and critique long chains of reasoning, supporting our rational thought (Mercier & Sperber 2017), it also triggers self-esteem reactions though FToM patterns. These patterns, in the absence of any correction from other people, are unreliable and irrational. Our sense of self, being based on unreliable inferences about 'what other people will think of me' is a second-hand, impoverished and unreliable view of ourselves – like seeing ourselves in a cracked mirror.

The human sense of self is linked to a sense of chronological time. This is a sense which most animals lack. For most animals, as far as we know they live in the moment. Past events in their lives are retained not as episodes, but only as learned regularities (e.g. where to find food), and they have no awareness of any distant future. They have no need for it. Only humans recall and arrange previous episodes in their lives, or consider a distant future. In part, this may have evolved through our need to converse, and to show off our intelligence by discussing past and future events; or merely to have those events as topics for conversation. Some of it is surely driven by self-esteem; one's perceived status in the group depends on how one thinks other people regard one's life - as a successful life, or not. We are constantly thinking about our own life history, past and future; polishing it to make it appear more successful in other peoples' eyes (as we think they see us).

Self-esteem is largely responsible for our view of ourselves as rational beings. An important part of our self-esteem is a set of ideas that: 'I am a rational person; I do not do things for no reason; I think things through'. So we often invent reasons, *post hoc*, for decisions we have taken for less rational reasons. We are less rational than we think we are (Chater 2022). Much of our reasoning is done not to decide what to do, but to explain why we have done it (Mercier & Sperber 2017), to others and to ourselves.

ToM patterns triggered by bodily feelings can lead to fast cascades, in which we first feel some emotion as a bodily feeling; then, using FToM patterns, we infer what our shadow audience will think of us if we show that emotion. This triggers further emotions, and further thoughts as we try to counter negative self-esteem. These fast cascades may drive the volatile, unpredictable, and irrational nature of human emotions. They are driven by FToM 'shadow audience' patterns which are learnt from an early age, and may never be un-learnt.

The need to impress other people causes group-think and tribalism. If some opinion is held within a group, and is affirmed in conversations, then we think we will achieve high status in the group by agreeing with it. We do this in our private thoughts, which are rehearsed conversations; self-esteem is enhanced by the inferred agreement of a shadow audience. It then matters more that some opinion should agree with an opinion of the group, than that it fits the facts and evidence. This is group-think.

Acceptance within one's own social group is enhanced by a negative view of other groups; an unfavourable comparison of 'them' with 'us'. This leads to tribalism and rejection of out-groups, reinforced by group-think. Tribalism need not appear extreme; it consists of any diminished or stereotyped view of people in another group – however that group may be defined (by appearance, wealth, belief, language, or education). In that sense, tribalism is universal. This diminished view of other people removes the need to think about what their life is actually like for them; it saves thinking about the impacts of one's own actions on them.

These are some of the irrational forces that cause people to mistreat and harm other people – man's inhumanity to man. This is no small matter – no mere scientific curiosity. Throughout history, in all parts of the world, groups of people have inflicted untold suffering on other people, and they continue to do so. Full language (and its irrationality) is the underlying reason why they do it.

## 9. Two Views of Language – Revisited

I have described how pragmatic conversational skills evolved by sexual selection for a prodigious language ability; how those skills require us to pre-consciously read other minds and to measure ourselves against the norms of the group.

With that understanding, we can re-visit the distinction between narrow language (syntax and semantics) and full language.

Narrow language is regarded, in a conventional view, as a neutral vehicle for rational thought, as shown in figure 3:

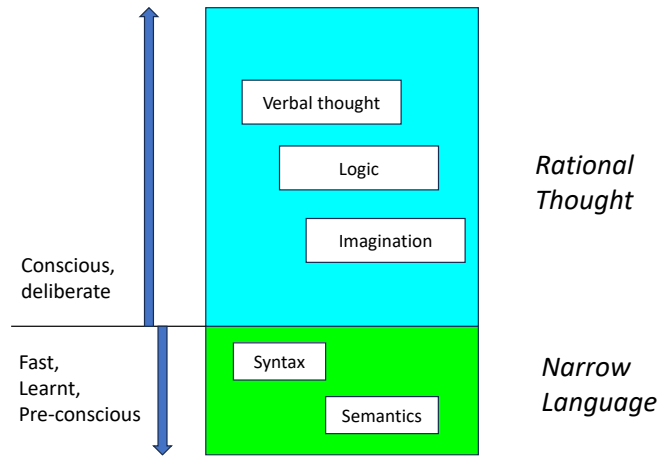

*Figure 3: A conventional view of language as a vehicle for rational thought.*

The line across the figure is the border between rational conscious thought and pre-conscious processes. The relative areas above and below the line seem to imply that the majority of our mental lives is rational. That is what we usually tell ourselves, and it fits our need for self-esteem.

Having seen the scope of full language, a more accurate picture of our mental lives is shown in figure 4:

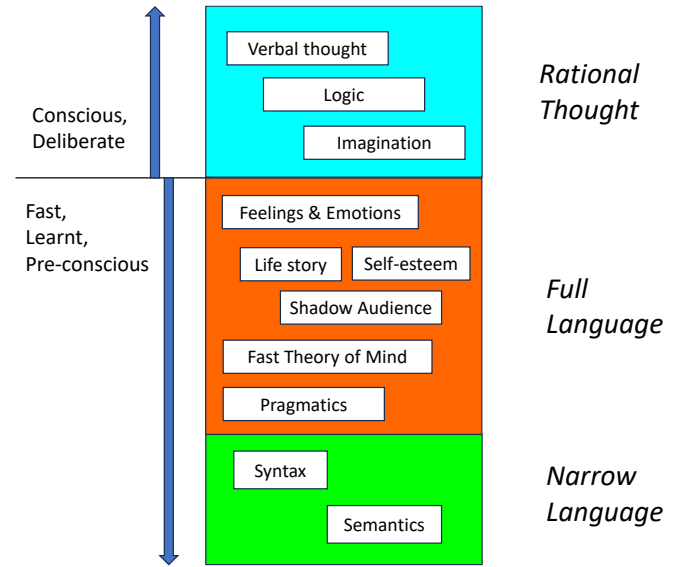

*Figure 4: The full role of language in our mental lives*

The horizontal line is still the borderline between the rational and the irrational. It shows that less of our mental lives is rational and deliberate; much more of it is fast, instinctive and pre-conscious. Language is not merely a neutral medium for expressing our rational thoughts. The central red region of full language is a control system for our social behaviour, and it acts faster than rational thought. It is centred on the shadow audience – a mental descendant of the conversational partners whom we need to impress, to get a mate. The shadow audience is always present behind our thoughts. It is as if our whole life was a play acted out in front of an audience we cannot see, whose thoughts and judgments we are trying to guess.

To compare the working of the shadow audience to narrow language:

- There is a process in our minds going from meanings or mental images, to word sounds or phrases. We know that this process (the green box) is a fast, pre-conscious way to create words and thoughts, but we think we can rationally critique the thoughts (blue box).
- There is a second process in our minds, going from thoughts or intentions, to an inferred response or judgment of ourselves by a shadow audience (this is the red box). It is also a fast, pre-conscious process; and sometimes inaccurate. It controls our motivations, feelings and actions, in ways we prefer not to admit.

## 10. Evolution of a Social Control System

The full language system in the human brain includes syntax and semantics, pragmatics, the Theory of Mind, our senses of self, biographical time, self-esteem and emotion. These are built on the fast computational mechanisms of feature structures and unification. Together, they control our social behaviour. For instance, when we feel a social emotion such as shame, that alters our behaviour. They enable us to form relationships with other people and to cooperate with them; to make extended trains of thought and plans – leading ultimately to all the benefits of civilisation. But they also lead to the ills of human society; such as how (in the pursuit of an illusory self-esteem) people are driven to attain power over other people, to harm them, and to curtail their freedoms.

In short, full language acts as a human social control system, controlling our social behaviour, for good or ill.

This social control system did not arise *de novo* with the evolution of human language. It has evolved from an earlier social control system of all primates.

In most mammalian species, reproductive success depends on physical fitness and strength – for instance, it is determined by threats and physical fights. In primates, this is not the case. Primate societies are complex, and individuals recognize relationships within their group, such as kinship and alliances. A primate with greater social intelligence can prevail by guile and alliances over physically stronger peers. The social intelligence which enables them to do this (Worden 1996a) involves learning and applying complex tree-like feature structures, which describe the social relationships and actions in a primate group.

The computational mechanisms of feature structures and unification, which underpin language, evolved in primates before the evolution of human language. Complex feature structures evolved with the first primates, up to twenty million years ago. It is then a small step to see that human language evolved from primate social intelligence (Worden 1996b).

Primate social intelligence evolved as a social control system, enabling primate social groups to function. Throughout the recent evolution of human language, it has remained a social control system – the use of feature structures and unification to control our social behaviour, by fast pre-conscious computations. Much of that behaviour is not uniquely human, novel, rational or free.

## 11. The Test of How Language Works

The ideas in this paper about how full language works can be checked by introspection and self-observation:

- When we hear any utterance, powerful pre-conscious pattern-matching processes take place in fractions of a second, before we are aware of its meaning.
- When we converse with somebody, we rapidly and pre-consciously infer their conversational intent
- In conversation, we are continually aware of what the other person might think about what we are going to say next.
- We infer what they might think about us, in the light of what we say.
- When contemplating any action, we work out 'what people will think of us' if we were to do it.
- Negative emotions follow events which negatively impact our self-esteem – events which affect how we think others perceive us.
- Many emotions are felt with greater intensity not when something happens, but when we realise that other people know it has happened.
- Emotions alter our intentions.
- Like a language, we learn emotional patterns in childhood, and they stay with us for all our lives.
- In any train of thought, we are aware of how a shadow audience in our minds will respond to that thought.
- We spend a lot of time constructing and rehearsing our own life story, worrying whether it is a successful life story – in the eyes of other people.
- We spend much of our mental lives comparing ourselves with other people, comparing their visible success with our own.
- We frequently form rapid, superficial and negative impressions of other people or groups.
- Negative assessments of other people or groups help to bolster our self-esteem.
- We frequently hold opinions in line with the opinions of our peers – confirming those opinions by the assumed agreement of our peers, rather than by examining the evidence.

(If any of these do not match your own experience, please let me know how!)

So our lives are less free than we suppose – running along the railway tracks of self-esteem and conformance with social groups. In this, we are directed by a pre-conscious language-based social control system.

## 12. Language, Freedom and Mindfulness

We spend most of our lives as social beings, guided by the social control system of language. Sometimes we are content for this to be; we are happy with how we behave, and other people are happy with it – or at least, they are not made unhappy by what we do. Sometimes our self-esteem can remain high, for valid reasons.

This is not always so. Either our self-esteem may be low, or it may be high for the wrong reasons. We may have treated other people as pawns in our game of self-esteem, and done them harm.

It sometimes appears that there is no escape from the social control system. When we try to think of a way out, we use language to do so; and language triggers the irrational pre-conscious mechanisms of self-esteem and emotion, which act faster than language. Rational thought is too slow, and cannot catch up with the irrational.

We spend the majority our lives as social beings, but we do not need to spend all our lives that way. It is possible to get off the social merry-go-round and just experience life. This happens to everyone, for some of the time; some of the most meaningful moments in life are the moments outside time, when one is so absorbed in the present moment that chronological time and self-esteem do not exist. This can happen in many ways – in physical activity, in music, in nature, in making something, in loving or giving. All these can be moments of freedom from the social self.

These moments can also be cultivated, in mindfulness and meditation. The core technique of meditation is to pay continued and close attention to feelings in the body, taking any sensation as an opportunity to examine it more closely (where is it? what is it like?) and not seeking any preferred or desirable feeling. This 'not seeking' in meditation is paradoxical, and each meditator learns their own way to approach it. The result can be that bodily feelings, which had previously been taken as codes for some social situation – as something to be avoided, or something to be dealt with – are no longer felt that way. They are just felt as they are, in the moment. This is a partial release from the social self.

The workings of full language may help to understand what happens in meditation. When feelings in the body are recognized for what they are – as conscious awareness of some region in space, which happens to be in the body – and are not taken as learnt codes for social situations, then the cascade of pre-conscious social reactions (a part of our language heritage) does not take place, or is diverted. We are then able to be simply aware of the present moment. Understanding how full language works can help us accept it, and not fight it, to diminish the cascades which sustain it. In this way, meditation can be understood as less of a mystery, and more as a simple release from a social control system which dominates our lives.

This release has beneficial effects, for the individual and for other people. It frees a person from forming simple, stereotyped caricatures of other people, using the Fast Theory of Mind. It allows a person to appreciate, slowly and more compassionately, what other people's lives are really like. Whether or not that makes them a 'better person' is in some sense the wrong question; to ask that question is to stay within the social control system, whereas the end point of mindfulness is to be free of the control system.

It has been said that 'Buddhism is like the sea. The sea is vast, and it has one taste – the taste of salt. Buddhism is vast, and it has one taste - the taste of freedom.' We can now see what that freedom is – it is freedom from the social control system of full language, which has had many baleful effects on human life. It is freedom from the central red region of figure 4. For most meditators it may be only a temporary freedom, but it enriches their lives.

If this paper can contribute in some small way to the scientific understanding of mindfulness, making it a bit less mysterious, then it may make mindfulness more approachable and accessible to more people.

## 13. AI and Large Language Models

There have recently been spectacular successes in the applications of artificial intelligence. Some of these, like Google DeepMind's AlphaFold, are significant scientific breakthroughs. The poster children of AI are Large Language Models (LLMs). How do LLMs compare with human language, as described in this paper?

Even when trained, and even in terms of narrow language (syntax and semantics) LLMs differ from narrow human language. Human narrow language is a two-way relation between words and meanings, where the meanings are related to the world in verifiable ways. The word 'ball' relates to round things, where the concept of roundness has a physical and geometric reality. Like Dr. Johnson, you can kick it. Large Language Models know none of this. The only thing they know about the word 'ball' is how it is used alongside other words in their vast training sets; so they can plausibly complete word patterns containing the word 'ball'. Their 'knowledge' is free-floating in the space of words, and has no contact with the world. Judged in terms of real-world knowledge, they frequently hallucinate.

Large Language Models are also in no sense a model for human language, because they require vast amounts of training data to learn a language – more training data than any person could experience in many lifetimes. They are also not a model of human language in the sense that (while nobody can describe how an LLM works, depending as it does on vast numbers of inscrutable neural connection weights), they certainly do not work by feature structures and unification – which are how human language is believed to work.

Some of these limitations of large language models may be surmountable; for instance, it is possible to embed a LLM in a robot with vision and limbs, so it can kick a ball, and start to relate its disembodied knowledge of words to knowledge of the world.

If LLMs can be enriched with real-world semantics, how might this extend from the semantics of narrow language to full language, as described in this paper? The capabilities of full language include a Fast Theory of Mind, reading contextual cues such as gesture and tone of voice, understanding of social norms, a concept of self and chronological time, and emotional responses to social situations mediated through bodily feelings. Today's LLMs have none of these.

Consider the Theory of Mind, or Dennett's (1989) Intentional Stance. When we converse with someone, we shape what we say (however imperfectly) to the needs of that person, as we understand their needs. As we converse, we build up a knowledge of what the other person thinks, knows, and wants, and that knowledge guides what we say.

LLMs do not do this in any explicit sense. LLMs have no internal representation of another person's mental states, and so do not update that representation in the light of their responses. It might be said that LLMs do this implicitly, by finding the most likely continuation for a conversation. But any human expectation that: 'The LLM understands what I know and what I want' is a misconception. The LLM does not understand you, because it has no concept of 'you'. It only knows how to continue a pattern of words.

As AI and robots move out of impersonal factory-like settings, towards more inter-personal applications like medical and social care, if LLMs are used for interaction and instruction, this issue will be critical. Typical current LLM scores on sandbox tests, like 70%, will not do. It will not be enough for an LLM to understand a user's intentions correctly 'most of the time', based on its impenetrable black-box learning from previous conversations; even low levels of errors by an LLM will be dangerous and will destroy human trust.

It is increasingly evident that for critical applications where errors are unacceptable, any LLM must be enclosed in a robust set of 'guardrails', built using real-world knowledge, and possibly, old-fashioned software engineering. An essential early step is to build in some of the knowledge of human full language – including a Theory of Mind to infer the knowledge and goals of the people it interacts with. As the human Theory of Mind is irrational and error-prone, this is a challenging task.

## 14. Conclusions

The human mind is partly rational, partly irrational. Our irrationality has caused immense harm over the ages, and continues to do so.

This paper has described the origins of human irrationality, in the sexual selection for language as a display of intelligence. The same fast pattern matching, which enables us to understand the words we hear, also drives our self-esteem and emotions, often in irrational and harmful ways.

If science is measured by its importance to mankind, then studying this is Big Science. It merits a worldwide cooperative effort, comparable to other fields of Big Science. It can lead to a scientific underpinning of history, of society, of individual lives, of human nature, and of human evil. The roots of human nature are to be found in full language – in self-esteem and the Theory of Mind. These are respectively how we misunderstand ourselves, and how we misunderstand other people. If, at some time in the future, we are to understand the sources of human evil as we now understand disease – and eventually cure human evil, as we now cure diseases – then understanding full language is the starting point.

## References

Bloom L et al. (1993) Language development from two to three, Cambridge University Press

Bybee J. (2010) Language, usage and cognition. Cambridge University Press

Chater, N. (2022) The Mind is Flat, Penguin, London

Christiansen M. H. and Kirby S. (Eds., 2003), Language evolution Oxford: Oxford University Press.

Chomsky N. (1965) Aspects of the theory of syntax. MIT Press, Cambridge, Mass.

Chomsky N. (1981) Lectures on government and binding. Foris.

Chomsky N. (1995) The minimalist program. MIT Press, Cambridge, Mass..

Croft, W. (2001). Radical construction grammar: Syntactic theory in typological perspective. Oxford University Press..

Dennett, D. (1989) The Intentional Stance, MIT press, Cambridge Mass.

Fillmore, C. (1985). Frames and the semantics of understanding. Quaderni di Semantica, 6, 222–254.

Fletcher P. and McWhinney B (1996) The handbook of child language, Blackwell, Cambridge, Mass.

Goldberg, A. E. (1995). Constructions: A Construction Grammar approach to argument structure. Chicago/London: University of Chicago Press.

Kaplan, R. M. and J. Bresnan (1981) Lexical Functional Grammar: a Formal System for Grammatical Representation

Kay P. (2002) An Informal Sketch of a Formal Architecture for Construction Grammar, Grammars 5: 1–19

Lande, R. (1981). Models of speciation by sexual selection on polygenic traits. Proc. Natl. Acad. Sci. U. S. A. 78, 3721–3725. doi: 10.1073/pnas.78.6.3721

Langacker, R. (2008) Cognitive grammar: A basic introduction. Oxford University Press.

Levinson, S. C. (1983). Pragmatics. Cambridge, UK: Cambridge University Press.

Levinson, S. C., and Torreira, F. (2015). Timing in turn taking and its implications for processing models of language. Front. Psychol. 6:731. doi: 10.3389/fpsyg.2015.00731

Maynard-Smith, J. (1982). Evolution and the theory of games. Cambridge, UK: Cambridge University Press.

Mercier, H., and Sperber, D. (2017). The enigma of reason. Harvard University Press.

Miller, G. (2001). The mating mind: How sexual choice shaped the evolution of human nature. London, Heineman.

Pinker S. (2021) Rationality, Allen Lane, London

Sag I. A., Boas H. C, & Kay P. (2012) Introducing Sign-Based Construction Grammar, in Sign-Based Construction Grammar, Eds: Boas H. C. & Sag I. A.

Sperber, D., and Wilson, D. (1986) "Relevance: communication and cognition" in Second edition (with postface) 1995 (Oxford: Blackwell)

Sperber, D., and Wilson, D. (2002) Pragmatics, modularity and mind-reading. Mind Lang. 17, 3–23. doi: 10.1111/1468-0017.00186

Stalnaker, R. (2002). Common ground. Linguist. Philosophy 25, 701–721. doi: 10.1023/A:1020867916902

Számadó, S. Z., and Szathmáry, E. (2006). Competing selective scenarios for the emergence of natural language. Trends Ecol. Evol. 21, 555–561. doi: 10.1016/j.tree.2006.06.02

Tomasello, M. (2014). A natural history of human thinking. Cambridge, Mass.: Harvard University Press.

Wittgenstein, L. (1958) Philosophical Investigations, second edition, Blackwell, Oxford

Worden, R.P. (1996a) Primate Social Intelligence, Cognitive Science 20, 579 - 616.

Worden, R. (1996b) The Evolution of Language from Primate Social Intelligence, in 'Approaches to the Evolution of Language' (Proceedings of the first EvoLang conference), Hurford et al, eds, Cambridge University Press

Worden, R. P. (1997) A Theory of language learning, https://arxiv.org/abs/2106.14612 .

Worden R. P. (2002) Linguistic Structure and the Evolution of Words, in 'Linguistic Evolution Through Language Acquisition', Briscoe E (ed.) Cambridge University Press

Worden, R. (2022a) The evolution of language by sexual selection, Front. Psychol., Vol 13 https://doi.org/10.3389/fpsyg.2022.1060510

Worden, R. (2022b) A Computational Model of Language Learning, unpublished paper and working computational model, at http://www.bayeslanguage.org/demo/model

Worden R (2024) The evolution of language and human rationality (short version of this paper), in Proceedings of the 14th biennial EvoLang conference on language evolution, Madison, Wisconsin